\documentclass[twocolumn,showpacs,prb]{revtex4}

\usepackage{graphicx}
\usepackage{dcolumn}
\usepackage{amsmath}
\usepackage{amssymb}

\makeatletter
\def\btt#1{\texttt{\@backslashchar#1}}%
\DeclareRobustCommand\bblash{\btt{\@backslashchar}}%
\makeatother

\begin{document}

\title[Short Title]{
Josephson interferometer in a ring topology as a symmetry prove of Sr$_2$RuO$_4$}

\author{Yasuhiro Asano}
\email{asano@eng.hokudai.ac.jp}
\affiliation{%
Department of Applied Physics, Hokkaido University, 
Sapporo 060-8628, Japan
}%

\author{Yukio Tanaka}
\affiliation{
Department of Applied Physics, Nagoya University, 
Nagoya 464-8603, Japan \\
CREST, Japan Science and Technology Corporation (JST), 
Nagoya 464-8603, Japan}%

\author{Manfred Sigrist}
\affiliation{
Theoretische Physik, ETH Honggerberg, 8093 Zurich, Switzerland
}%

\author{Satoshi Kashiwaya}
\affiliation{
National Institute of Advanced Industrial Science and Technology, 
Tsukuba 305-8568, Japan
}%

\date{\today}

\begin{abstract}
The Josephson effect is theoretically studied in two types of SQUIDs
consisting  
of $s$ wave superconductor and Sr$_2$RuO$_4$. Results show various
response of  
the critical 
Josephson current to applied magnetic fields depending on the type of SQUID 
and on the pairing symmetries. 
In the case of a $p_x+ip_y$ wave symmetry, 
the critical current in a corner SQUID becomes an asymmetric function
of magnetic  
fields near the critical temperatures. 
Our results well explain a recent experimental finding 
[Nelson et. al, Science \textbf{306}, 1151 (2004)]. 
We also discuss effects of chiral domains on 
the critical current.
\end{abstract}

\pacs{74.50.+r, 74.25.Fy,74.70.Tx}
\maketitle

\section{introduction}

One of the important developments for unconventional superconductivity 
after the the discovery of high-$T_c$ superconductivity~\cite{bednorz}, has been the series
of so-called phase sensitive experiments which test for the symmetry of the Cooper pair wave function.
These are the SQUID-type of interference
experiments~\cite{wollman,sigrist1,ott,mathai}, the observation of 
spontaneous half-flux quantization in frustrated loops~\cite{tsuei} and the measurement of
zero-bias peaks in quasiparticle tunneling spectra indicating subgap quasiparticle states at the
sample surface \cite{hu,yt95l,geerk,ya04-1,RPP}. This set of experiments,
technically rather diverse, is based on the same concept, the unconventional phase structure
of the superconducting condensate, and have uniquely proven that the Cooper pairs have
the spin-singlet $d_{x^2-y^2}$ wave symmetry.  A further unconventional superconductor
whose pairing symmetry has been established with high confidence is Sr$_2$RuO$_4$~\cite{maeno}.
This is
a spin triplet $p$-wave superconductor~\cite{ishida,duffy,luke} with a
gap function of the form $ \boldsymbol{d}(\boldsymbol{k}) \propto
\hat{\boldsymbol{z}} (p_x \pm i p_y ) $ \cite{rice,mackenzie}, a
so-called chiral $p$-wave state with whose Cooper pairs possess an
angular momentum component along the $z$-axis.  
Tunneling experiments have shown the presence of subgap surface states \cite{laube,mao,kawamura}.
Moreover, the anomalous temperature dependence of the critical current in Pb/Sr$_2$RuO$_4$/Pb
Josephson junction arrangement \cite{jin} has been interpreted in terms of an interference
effect~\cite{honer,yamash,yamashiro}. A direct experiment of the type
of a SQUID-interference or frustrated loop
is difficult here for many technical reasons and has not been performed until very recently~\cite{nelson}. 

In this paper we would like to analyze some issues which have to be
taken into account in the interpretation of the SQUID-type experiments
for Sr$_2$RuO$_4$. 
The basic principle had been designed long ago by Geshkenbein and co-workers~\cite{gesh}. 
One of the problems lies in the Josephson junctions between a conventional 
spin-singlet $s$-wave superconductor and a spin-triplet $p$-wave superconductor. The mismatch of
the angular moment (parity) and of spin quantum number on the two sides of a Josephson junction
seems at first sight inhibit the lowest order Josephson effect so that only a coupling in second order
would be allowed. This would indeed be fatal for phase sensitive tests
based on the Josephson effect.  
It has, however, been shown that
the presence of spin-orbit coupling saves the situation 
since only the total angular momentum has to be conserved. 
Thus under well-defined conditions, the lowest order between an $s$-
and a $p$-wave superconductor is
possible~\cite{geshkenbein,millis,sigrist2,ya01-3,ya03-1}. The conditions
leave a certain arbitrariness concerning the sign of the Josephson
coupling which can be important for interference effects. This 
can be illustrated, if we consider the definition of the lowest order
matrix element which can be  
derived from a simple microscopic tunneling model:
\begin{equation}
\langle \psi_s (\boldsymbol{k}) (\boldsymbol{k} \times \boldsymbol{n} ) \cdot \boldsymbol{d}(\boldsymbol{k}) \rangle_{FS}
\label{basic}
\end{equation}
where the average runs over the Fermi surface and $ \boldsymbol{n} $ is the interface normal vector. 
The arbitrariness appears in the orientation of the normal vector,
into or out of the $p$-wave superconductor. This has been recently
demonstrated by Asano and co-workers~\cite{ya03-1} 
 for a model where the 
spin-orbit coupling of the interface potential was taken into account for this matrix element whose
sign then depends on the shape of the interface potential. Thus details of the spin-orbit coupling and
the interface potential, e.g. the way the parity is broken at the interface, 
matter for the Josephson phase relation. The arrangement suggested by
Geshkenbein et al. relies on the assumption that all interfaces treat
parity the same way \cite{gesh}. We will follow this assumption here
too.   

We discuss now two basic forms of SQUID-interference devices built
from an $s$- and a $p$-wave superconductor. The first type in
Fig.~\ref{fig1}(a) is 
the symmetric SQUID as proposed by Geshkenbein and co-workers \cite{gesh} and the second
in Fig.~\ref{fig1}(b) is the corner-SQUID analogous to the one used for high-$T_c$-superconductors. 
The $ p_x + i p_y $-phase of Sr$_2$RuO$_4$ introduces an additional problem. This phase is
two-fold degenerate, i.e. $ p_x + i p_y $ and $ p_x - i p_y $ are equivalent and would in general
form domains in a sample, depending on the cooling history. We will discuss in the following also the
implications domains on the interference experiment. Similar problems
appear for other chiral superconducting phase as we will discuss
below.  

This paper is organized as follows. 
In Sec.~II, the Josephson effect in two types of SQUID
is discussed for $p_x+ip_y$, $p_x$ and $p_y$ pairing symmetries. 
Effects of the chiral domain are studied in Sec.~III. 
In Sec.~IV, we discuss the critical current in the chiral $d$ and the chiral $f$
wave symmetries. The discussion and the conclusion are given in Sec.~V and XI, 
respectively.

\section{josephson current}
\begin{figure}[htbp]
\begin{center}
\includegraphics[width=8.0cm]{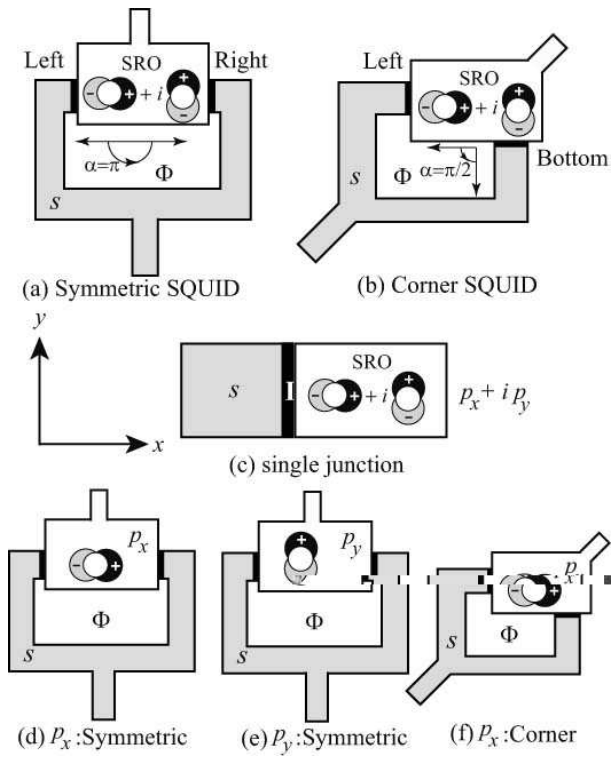}
\end{center}
\caption{
Schematic pictures of two types Josephson junction are shown in (a) and (b). 
The singly connected Josephson junction is given in (c).
In (d)-(f), the SQUID of $p_x$ and $p_y$ symmetries are shown. The Josephson effect in 
the corner SQUID of $p_y$ symmetry is identical to that
of the $p_x$ symmetry.
}
\label{fig1}
\end{figure}

Before discussing the SQUID-experiment, the basics of the Josephson current-phase relation
between $s$ wave superconductor and Sr$_2$RuO$_4$ (SRO) in Fig.~\ref{fig1}(c) has to be addressed. 
We describe the gap function of Sr$_2$RuO$_4$ by~\cite{rice,barash}
\begin{align}
\hat{\Delta}_{\boldsymbol{p}}=& i ( \eta_x \bar{p}_x + \eta_y \bar{p}_y )\hat{\boldsymbol{z}}\cdot \hat{\boldsymbol{\sigma}} \hat{\sigma}_2  \\
=&  i\Delta ( \bar{p}_x \pm i \bar{p}_y ) 
\hat{\boldsymbol{z}}\cdot \hat{\boldsymbol{\sigma}} \hat{\sigma}_2 \\
=& i \Delta e^{i \theta} \hat{\boldsymbol{z}}\cdot \hat{\boldsymbol{\sigma}} \hat{\sigma}_2,
\end{align}
where $\hat{\sigma}_j$ with $j=1,2$ and 3 are the Pauli matrices, $ \eta_x $ and $ \eta_y $ are
the two complex order parameters and 
$\hat{\boldsymbol{z}}$ is taken as a unit vector parallel to the $ z $-axis.\cite{mackenzie}
Moreover,  $\bar{p}_x=p_x/p_F=\cos\theta$ ($\bar{p}_y=p_y/p_F=\sin\theta$) is the 
normalized momentum
component on the Fermi surface in the $x$ ($y$) direction with $p_F$ being the 
Fermi momentum of the
$p$-wave superconductor.  Assuming a cylindrical symmetric Fermi surface, we can represent this
gap function also simply to the angle $ \theta $ on the Fermi surface, reflecting best its internal
phase structure. 
The gap function of the $s$-wave superconductor is given by
$\hat{\Delta}_{\boldsymbol{k}}= i \Delta \hat{\sigma}_2$. Without any loss of generality, we may
take the gap magnitudes identical in both superconductors, $ \Delta $. 
On the basis of the current-phase relation~\cite{millis,sigrist2,ya01-3}, the Josephson
current for the lowest two orders derived from a microscopic calculation close to $ T_c $~\cite{ya03-1} 
can be written as 
\begin{align}
\bar{J}_{p_x\pm ip_y}=& 
J_1 \cos(\varphi + \theta_n) - J_2 \sin\left(2(\varphi+\theta_n)\right), \label{pxy1}\\
J_1=&T_B \;  \alpha_S \left(\frac{\Delta}{2T}\right)
\frac{\Delta}{\Delta_0},\label{j1}\\
J_2=&T_B^2\left(\frac{\Delta}{2T}\right)^3\label{j2}\frac{\Delta}{\Delta_0},
\end{align}
where $ \theta_n $ is the angle of the junction normal vector in the plane relative to 
the $x$-axis, 
$ T_B $ denotes the transmission probability of a Cooper pair and 
$\alpha_S $ is a measure for the strength of the spin orbit coupling. Note that it can have
either sign depending on the junction.  
The Josephson current $\bar{J}$ is measured in units of $e\Delta_0N_c/\hbar$,
where $\Delta_0$ is the amplitude of the gap function at $T=0$ and $N_c$ is the 
number of propagating channels on the Fermi surface.
Constant coefficients of the order of unity have been omitted in Eq.~(\ref{pxy1}). 
From Eq.~(\ref{basic}) it is clear that the coupling between $ s$-wave and $ p_x + i p_y $-wave
superconductor for a junction in $x$-direction ($\theta_n =0 $) 
goes via the $ p_y $ component. 
We assume a gauge where the overall phase of the $ p $-wave 
order parameter is shifted by $ \pm \pi/2 $ for the $ p_y $-component. Thus, in lowest order
the current-phase relation is 
proportional to $\cos\varphi$ rather than $ \sin \varphi $, in Eq.~(\ref{pxy1}). 

In the first step, 
we consider the interference pattern of the two basic SQUID's in Fig.~\ref{fig1}(a) 
and (b). We assume that all junctions are of the same in the sense, that the normal vector $ \boldsymbol{n} $ in Eq.~(\ref{basic}) is pointing a definite direction, say towards the $s$-wave superconductor. 
In the symmetric SQUID, the Josephson current of the left and right junction are given by
\begin{align}
J_L(\varphi) =& J_1 \cos\varphi - J_2 \sin 2\varphi,\\
J_R(\varphi) =& -J_1 \cos\varphi - J_2 \sin 2\varphi.
\label{sym}
\end{align}
The relative sign change between the first-order terms on both sides is due to the 
the corresponding $ \theta_n $ which differ by $ \pi $. 
The Josephson current in the symmetric SQUID is then expressed by~\cite{barone}
\begin{align}
J_{\textrm{S}}(\varphi,\Phi)=& J_L(\varphi+\phi_B) + J_R(\varphi-\phi_B),\\
=& -2J_1 \sin\varphi\sin\phi_B - 2 J_2 \sin 2\varphi \cos 2\phi_B,
\end{align}
where $\phi_B= \pi {\Phi}/{\Phi_0}$ and $\Phi_0 = {2\pi\hbar c}/{e}$.
In the corner SQUID, the Josephson current in the bottom
junctions is given by
\begin{align}
J_B(\varphi) = -J_1 \sin\varphi + J_2 \sin 2\varphi,
\label{corner}
\end{align}
where the current-phase relation is derived from Eq.~(\ref{pxy1}) with 
$\theta_n=\pi/2$. 
The Josephson current in the corner SQUID is expressed in the same way,
\begin{align}
J_{\textrm{C}}(\varphi,\Phi)
=& J_1 \cos(\varphi+\phi_B) - J_1\sin(\varphi-\phi_B) \nonumber \\ 
&- 2 J_2 \cos 2\varphi \sin 2\phi_B.
\end{align}
Note that for the corner SQUID it matters which state is realized. For Eq.~(\ref{corner}) we 
assumed that the state $ p_x + i p_y $, while for $ p_x - i p_y $ would yield a sign change of the first term.

\begin{figure}[htbp]
\begin{center}
\includegraphics[width=8.0cm]{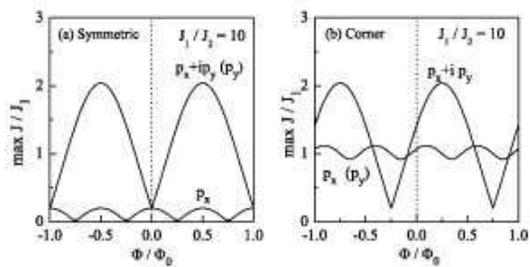}
\end{center}
\vspace{-0.8cm}
\caption{
The Josephson critical current is plotted as a function of $\Phi$ in the symmetric
SQUID in (a) with $J_1/J_2=10$.
The calculated results for the corner SQUID are shown in (b).
}
\label{fig2}
\end{figure}
In Fig.~\ref{fig2}, the critical Josephson current in the symmetric SQUID and 
that in the corner SQUID are shown as a function of $\Phi$.
A parameter $J_1/J_2=10$ is realized in high temperatures near $T_c$. 
Since $J_1 \gg J_2$, we find,
\begin{align}
\textrm{max} |J_{\textrm{S}}(\Phi)| \simeq& 2J_1 
\left| \sin\left(\pi \frac{\Phi}{\Phi_0}\right)\right|, \label{jsht}\\
\textrm{max} |J_{\textrm{C}}(\Phi)| \simeq& 2J_1 
\left| \sin\left(\pi \frac{\Phi}{\Phi_0}\pm\frac{\pi}{4}\right)\right|. \label{jcht}
\end{align}
The odd parity symmetry immediately results in a minimum of the critical current 
at $\Phi=0$ in the symmetric SQUID~\cite{gesh}, which is connected with the
opposite sign of Josephson coupling on the two junctions according to Eq.~(\ref{sym}).
Note that this pattern is not dependent on which of the two degenerate $p$-wave state is realized. 
Actually for the symmetric SQUID any $p$-wave pairing state gives the same qualitative
behavior and does not depend on the detailed symmetry, such as broken
time reversal symmetry, as long as the first-order Josephson coupling
induced by spin-orbit effects is dominant. 
This is different for the corner SQUID which leads to a distinction between different $p$-wave
states. 

The change to the corner SQUID configuration yields a phase shift by $ \pm \pi/2 $ for 
 $p_x \pm ip_y$.  As a consequence, the $\cos\varphi$ current-phase relation can be exchanged
by   $\sin\varphi$ for the bottom junction. 
For comparison, the results for the $p_x$- and $p_y$-wave states are depicted in Fig.~\ref{fig2}
with the corresponding device illustrations in Fig.~\ref{fig1}(d)-(f). 
The Josephson current-phase relation in the single junctions can for fixed gauge be given 
by~\cite{ya03-1,note}
\begin{align}
J_{p_x}(\varphi)=&-J_2\sin 2\varphi,\label{px}\\ 
J_{p_y}(\varphi)=& J_1 \sin\varphi - J_2\sin2\varphi.\label{py}
\end{align}
For the $p_x$-component ($ \boldsymbol{n} $ parallel to $x$), there is no contribution due to
spin-orbit coupling, in contrast to the $ p_y $-component which has such a term proportional to $\sin\varphi$. in the $p_y$ symmetry. 

In the symmetric SQUID of the $p_x$-type, the critical current has the  period 
of $\Phi_0/2$ as shown in Fig.~\ref{fig2}(a) because there is only the second-order contribution
in the Josephson coupling. On the other hand, 
the $p_y$-type behaves  identical to the $p_x+ip_y$-type.
For the corner SQUID configuration the $p_x$-symmetry (and equivalently for the $p_y$-symmetry)
 has again a $\Phi_0/2$-periodic
interference pattern of the critical current.

\begin{figure}[htbp]
\begin{center}
\includegraphics[width=8.0cm]{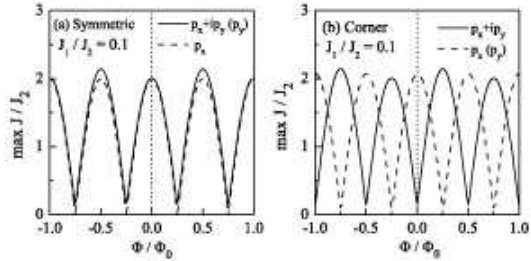}
\end{center}
\vspace{-0.8cm}
\caption{
The Josephson critical current is plotted as a function of $\Phi$ for $J_1/J_2=0.1$.
}
\label{fig3}
\end{figure}

If we assume that $J_1 \ll J_2$ due to weak spin-orbit coupling, the features of
the interference pattern are significantly modified.
In Fig.~\ref{fig3}, the critical Josephson current is plotted as a function of $\Phi$
for $J_1=0.1J_2$. 
For $J_1 \ll J_2$, we actually find
\begin{align}
\textrm{max} |J_{\textrm{S}}(\Phi)| \simeq& 2J_2 
\left| \cos\left(2\pi \frac{\Phi}{\Phi_0}\right)\right|, \label{jnlt}\\
\textrm{max} |J_{\textrm{C}}(\Phi)| \simeq& 2J_2 
\left| \sin\left(2\pi \frac{\Phi}{\Phi_0}\right)\right|, 
\end{align}
for the $p_x+ip_y$ symmetry.
The period of $J_c$ is given by $\Phi_0/2$ because the Josephson current proportional
to $\sin 2\varphi$ is dominant. 
The critical current in the symmetric SQUID takes its maxima at $\Phi=0$
because $\sin 2\varphi$ remains unchanged 
under the $\pi$ phase shift.
On the other hand in the corner SQUID,
the critical current takes its minima at $\Phi=0$.
This is due to the sign change of $\sin 2\varphi$
under the $\pi/2$ phase shift in the $p_x+ip_y$ symmetry.
The critical current in the $p_x$ and $p_y$ symmetries 
can be described by Eq.~(\ref{jnlt}) irrespective of types of SQUID
 as shown in the Fig.~\ref{fig3}. 
Thus the minima at $\Phi=0$ in the corner SQUID
directly suggests the $p_x+ip_y$ symmetry in SRO when 
the period of the oscillations is
$\Phi_0/2$.

\begin{figure}[htbp]
\begin{center}
\includegraphics[width=8.0cm]{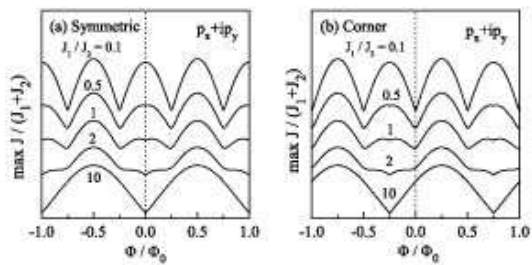}
\end{center}
\caption{
The Josephson critical current is plotted as a function of $\Phi$ for $p_x+ip_y$.
}
\label{fig4}
\end{figure}
In Fig.~\ref{fig4}, the critical current for the $p_x+ip_y$ symmetry 
is shown for several choices of $J_1/J_2$.
In the vertical axis, successive plots have been offset.
The critical current is always symmetric with respect to $\Phi$ in the symmetric SQUID.
The current maxima at $\Phi_0$ for $J_1/J_2=0.1$ is changed to the minima as increase of
$J_1/J_2$.
In the corner SQUID, the asymmetry in the critical current gradually disappears
with decreasing of $J_1/J_2$. 

\section{chiral domains}
\begin{figure}[htbp]
\begin{center}
\includegraphics[width=8.0cm]{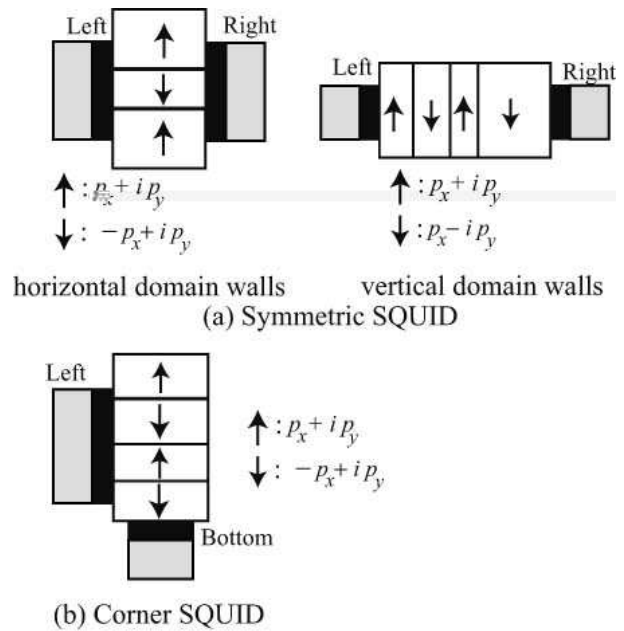}
\end{center}
\vspace{-0.8cm}
\caption{
 A simple model of chiral domains is illustrated in the symmetric SQUID (a) and 
the corner SQUID (b).Across the domain wall with a normal vector $\boldsymbol{n}$,
the $p$-wave component parallel to $\boldsymbol{n}$ keeps its phase, while the component
perpendicular to $\boldsymbol{n}$ changes the phase by $ \pi $.}
\label{fig5}
\end{figure}

We have assumed so far that SRO is a single domain of the $p_x+ip_y$ symmetry. 
Real materials, however, may have multi-domain structures of the $p_x+ip_y$
and the $p_x-ip_y$ symmetries.
Here we discuss effects of chiral domains on the Josephson current.
We consider a simple model of such chiral domains as shown in 
Fig.~\ref{fig5}, where $\uparrow$ and $\downarrow$ indicate domains
of the different chiral states.
The size of domains should be much larger than the thickness of domain walls
which is typically given by the coherence length of superconductors. 
The structure of domain walls has been investigated based on Ginzburg-Landau 
theories~\cite{sigrist3}. For a domain wall with normal vector $ \boldsymbol{n} $, 
the $ p $-wave
component parallel to $\boldsymbol{n} $, $ p_{\parallel} $, 
keeps its phase, while the component
perpendicular, $ p_{\perp} $, changes the phase by $ \pi $.  This has important implication to
the analysis of the above SQUID configurations which we will discuss here for a few 
simple situations as shown in Fig.~\ref{fig5}, under the assumption $
J_1 \gg J_2$. 

\begin{figure}[htbp]
\begin{center}
\includegraphics[width=8.0cm]{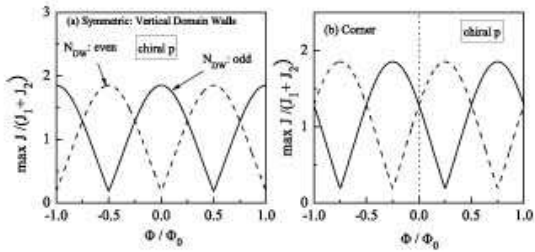}
\end{center}
\vspace{-0.8cm}
\caption{
 The critical current in the presence of domain walls for $J_1/J_2=10$.
In the symmetric SQUID, the results for the chiral $p$ wave symmetry 
in the presence of the vertical domain walls are shown in (a).
In (b), the critical current for the chiral $p$ wave symmetry
 in the corner SQUID is shown. The solid (broken) line 
 denote the results when the chirality of the last domain at the bottom
is $p_x+ip_y$ ($-p_x+ip_y$) in (b). 
}
\label{fig6}
\end{figure}

The presence of {\it horizontal} domain walls in Fig. \ref{fig5}, i.e. 
$ \boldsymbol{n} \parallel y $, implies that
the two chiral domains have either $ p_x + i  p_y $ or $ - p_x + i p_y $ form. According
to our previous discussion, no qualitative change of the interference pattern is 
expected. Thus the critical current in Fig.~2(a) remains unchanged even in the 
presence of the horizontal domain walls in the symmetric SQUID.
On the other hand, the vertical domain walls give rise to domains with $p_x +i p_y$ and
$p_x - i p_y$ form. In this case the number for domain walls between the two junctions
matters since each domain wall switches the phase of $ p_y$, entering
the first-order term of the Josephson coupling, by $\pi$. Thus, the
interference pattern would look different for an odd or 
even number of domain walls:
\begin{align}
\textrm{max} |J_{\textrm{S}}(\Phi)| \simeq \left\{
\begin{array}{cc}
 2J_1 
\left| \sin\left(\pi \frac{\Phi}{\Phi_0}\right)\right| & N_{DW}\;  \text{even}\\
 & \\
 2J_1 
\left| \cos\left(\pi \frac{\Phi}{\Phi_0}\right)\right| & N_{DW}\; \text{odd},
\end{array}\right.
\end{align}
where $N_{DW}$ is the number of the vertical domain walls. 
If the number of walls is an even integer,
the critical current takes its minimum at $\Phi=0$ as shown in Fig.~\ref{fig6}(a).
However, the critical current has its maximum at $\Phi=0$
when the number of domain walls is an odd integer. 
Thus the existence of the vertical domain walls can change drastically the
characteristic feature of the critical current in the symmetric SQUID.

In the corner SQUID, the current-phase relation in the two junctions as shown in 
Fig.~\ref{fig5}(b) are given by
\begin{align}
J_L(\varphi) =& J_1 \cos\varphi - J_2 \sin 2\varphi,\\
J_B(\varphi) =& \pm J_1 \sin\varphi + J_2 \sin 2\varphi.
\end{align}
where the sign of the first-order term of the bottom junction is determined by
the chirality of the last domain at the bottom. The number of domain walls is irrelevant 
here.
Still the asymmetry of the interference pattern indicates the broken time reversal 
symmetry
in any case as shown in Fig.~\ref{fig6}(b).

\section{Other chiral superconducting phases}
The chirality of superconductivity reflects 
the internal angular momenta of a Cooper pair. 
Also other superconducting phases besides the chiral $ p $-wave phase can be found
with this property. 
In this section, we compare the Josephson effect of the $p_x+ip_y$-wave phase
with those of other chiral phase of  $d$- and $f$-wave origin.

A chiral phase in the case of spin-singlet 
$d$-wave symmetry can be composed of the $ d_{x^2-y^2} $ and the $ d_{xy} $-state, when they 
are degenerate, yielding  a $ d_{x^2-y^2} \pm i d_{xy}$.
This gap function was discussed as the surface states of high-$T_c$ 
cuparates~\cite{laughlin} where the degeneracy is not given, but one of the
components is subdominant. The other example is 
Na$_x$CoO$_2\cdot y$H$_2$O~\cite{ogata}.  
Thus we consider the gap function 
\begin{align}
\hat{\Delta}_{\boldsymbol{k}}^{(d)}=&  \left\{ \eta_x (\bar{p}_x^2-\bar{p}_y^2) + \eta_y  2 \bar{p}_x \bar{p}_y \right\} i \hat{\sigma}_2  \\ 
=& 
\Delta (\bar{p}_x^2-\bar{p}_y^2 \pm i 2 \bar{p}_x \bar{p}_y) i \hat{\sigma}_2 
=  \Delta e^{\pm i2 \theta} i \hat{\sigma}_2.
\end{align}
Here the coupling to the $s$-wave superconductor
is simpler, since it is not relying on spin-orbit coupling. Thus, 
the current phase relation dominated by the first-order coupling. 
\begin{align}
J (\phi) = J_d \sin (\phi + 2 \theta_n), \\
J_d = T_B \left( \frac{\Delta}{2T} \right) \left( \frac{\Delta}{\Delta_0} \right).
\end{align}

Concerning the domain wall structure of this state, the two order parameter components
behave in the same way as in the $ p$-wave case. Thus only the component perpendicular
to the domain wall normal vector changes sign. Thus, for $ \boldsymbol{n} \parallel x $ the
component $ \eta_y $ ($d_{xy} $) changes sign while $ d_{x^2 - y^2} $ keeps the phase. 
Thus, irrespective of the domain wall number for the vertical domain wall case, the maximum
of the critical current in the symmetric SQUID would be at $ \Phi= 0 $. Thus, domain walls
could not lead to an interference pattern like in the $p$-wave case. In the case of horizontal
domain walls, the situation is more complicated. 
There is a staggered phase for the $d_{x^2-y^2} $-
component. This would give rise to compensating contributions of the two domains to the Josephson effect, similar to the situation discussed in the context of $c$-axis junction between an $s$-wave
and a twinned high-temperature superconductor\cite{SKMRL,Agter}. In addition, this configuration
could give rise to spontaneous half-quanta flux lines 
wherever a domain wall hits the junction perpendicularly.  
Under these circumstances, the interpretation of interference pattern would require much more
care. Note that such spontaneous fluxes do not occur in the $p$-wave case as long as
the domain wall ends perpendicularly on the junction. 
In both cases, the flux magnitude depends on the 
angle between junction interface and domain wall.
A similar problem with the domain walls occurs for the corner junction 
as a simple examination reveals. 

\begin{figure}[htbp]
\begin{center}
\includegraphics[width=8.0cm]{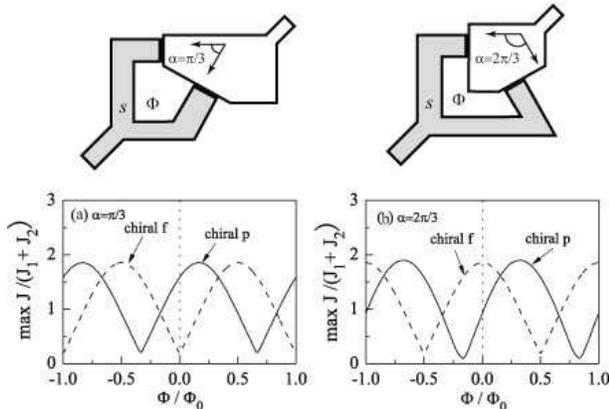}
\end{center}
\vspace{-0.8cm}
\caption{
The critical current for the chiral $p$ wave is compared with 
that for the chiral $f$ wave in (a) and (b), where $J_1/J_2=10$.
The relative junction angle is set at $\alpha=\pi/3$ in (a) and $\alpha=2\pi/3$
in (b) as shown in the upper figures.
}
\label{fig7}
\end{figure}

The spin-triplet chiral $f$-wave symmetry is also a possible candidate for 
the superconducting phase in Na$_x$CoO$_2\cdot y$H$_2$O. 
The proposed gap function is given by
\begin{align}
\hat{\Delta}_{\boldsymbol{k}}^{(f)}=& i\Delta 
\left\{ (\bar{p}_x^2-3\bar{p}_y^2)\bar{p}_x  \pm i (\bar{p}_y^2-3\bar{p}_x^2)\bar{p}_y \right\} 
 \hat{\boldsymbol{z}}\cdot \hat{\boldsymbol{\sigma}} \hat{\sigma}_2 \nonumber \\
=&  \Delta e^{\mp i 3 \theta}  i \hat{\boldsymbol{z}}\cdot \hat{\boldsymbol{\sigma}} 
\hat{\sigma}_2.
\end{align}
The current phase relation for the chiral $f$ wave symmetry is given by
Eq.~(\ref{pxy1}).
In both the symmetric and the corner SQUID, the characteristic behavior of the 
critical current for the chiral $p$ wave symmetry discussed in 
Sec.~II and III are also valid for the chiral $f$ wave symmetry.
 In the limit of $J_1/J_2\gg 1$, the critical current in the symmetric 
SQUID take the minima at $\Phi=0$
and that in the corner SQUID shows the phase shift by $\pi/4$
in the absence of the domain walls. 
In addition, there are no differences in effects of domain walls on the critical
current between the chiral $p$ wave and the chiral $f$ one.
To distinguish the chiral $p$ wave symmetry from the chiral $f$ wave, 
we should design SQUID's with $\alpha=\pi/3$ or $2\pi/3$~\cite{sauls}. 
The critical current for $f$ waves becomes symmetric function of $\Phi$
and takes its minimum (maximum) at $\Phi=0$ in the SQUID with 
$\alpha=\pi/3$ ($\alpha=2\pi/3$)~\cite{sauls} as shown in Fig.~\ref{fig7}. 
On the other hand, the critical current for $J_1/J_2\gg 1$ in the chiral 
$p$ wave results in
\begin{align}
\textrm{max}\left|J_{\alpha=\pi/3}(\Phi)\right| &\simeq 2J_1 
\left|\cos\left(\pi\frac{\Phi}{\Phi_0}-\frac{\pi}{6}\right)\right|,\\
\textrm{max}\left|J_{\alpha=2\pi/3}(\Phi)\right| &\simeq 2J_1 
\left|\cos\left(\pi\frac{\Phi}{\Phi_0}-\frac{\pi}{3}\right)\right|.
\end{align}
The calculated results are shown in Fig.~\ref{fig7}.
The asymmetry in the critical current 
persists in the chiral $p$ wave in such SQUID's. 
This argument, however, is valid when
the difference in the transmission probabilities of the two tunnel 
junctions in Fig.~\ref{fig7} are much smaller than themselves.

\section{discussion}
We have shown that the asymmetry of the critical current with respect to $\Phi$
in the corner SQUID is the evidence of the $p_x+ip_y$ symmetry
when the period of oscillations is $\Phi_0$.
A recent experiment shows a large asymmetry of the critical current in the corner 
SQUID~\cite{nelson}.
The experiment also shows a minima of the critical current around $\Phi=0$ 
in the symmetric SQUID.
The position of minima, however, slightly deviates from $\Phi=0$.
We have assumed that $J_1$ in the left junction is equal to that 
in the right junction. When those are different from each other, the Josephson
critical current becomes asymmetric of $\Phi$ even in the symmetric SQUID. 
Thus the small asymmetry found in the experiment might be caused by the 
asymmetry of the two tunnel junctions. 
The degree of asymmetry is negligible when the difference of $J_1$ 
in the two junctions is much smaller than themselves.

In addition to $p_x+ip_y$, the $\sin(p_x) +i \sin(p_y)$~\cite{miyake,kuwabara}
and $\sin(p_x+p_y)+i\sin(p_x-p_y)$ symmetries~\cite{arita} 
have been proposed so far. The perturbation
expansion~\cite{nomura,yanase} also indicates a possibility of 
the chiral $p$ wave symmetry with more complicated structure than
$\sin(p_x) +i \sin(p_y)$ and $\sin(p_x+p_y)+i\sin(p_x-p_y)$.
The characteristic behavior of the Josephson current in the two types of
SQUID discussed in Sec.~II and III
are valid not only for the isotropic $p_x+ip_y$ symmetry but also for
anisotropic $\sin (p_x)+i\sin (p_y)$ and 
$\sin(p_x+p_y)+i\sin(p_x-p_y)$ symmetries.
The latter two gap function have the four-hold symmetry in the 
momentum space consistently with a thermal conductivity experiment~\cite{izawa}.
In particular, $\sin(p_x+p_y)+i\sin(p_x-p_y)$ is close to the chiral $f$
wave symmetry because $\sin(p_x+p_y)$ and $\sin(p_x-p_y)$ change their sign
six times on the Fermi surface. 
To distinguish one symmetry from another, much more detail analysis
would be required.
At present, we propose the angle resolved tunneling spectra for this 
analysis~\cite{tamura}.

\section{conclusion}
We have studied the critical Josephson current in two types of
SQUID consisting of $s$ wave superconductor and Sr$_2$RuO$_4$. 
In the $p_x+ip_y$ symmetry, the critical Josephson current in the corner 
SQUID becomes the asymmetric function of $\Phi$ because 
 the current-phase relation in the two 
junctions relate to each other by $\pi/2$ phase shift in the gap function.
We also show that the asymmetry remains even in the presence of the chiral
 domain structures in Sr$_2$RuO$_4$.
Our results well explain the recent experimental 
findings~\cite{nelson}.

\begin{acknowledgments}
The authors thank Y.~Maeno, Y.~Liu  and V.B. Geshkenbein for useful discussion.
This work has been partially supported by Grant-in-Aid for 
the 21st Century 
COE program on "Topological Science and Technology" and "Frontiers of Computational Science" 
from the Ministry of 
Education, Culture, Sport, Science and Technology of Japan.
\end{acknowledgments}

{} 


\begin{thebibliography}{}
\bibitem{note} Here we assume temperatures are still higher than $\Delta_0T_B$.
When $T\ll\Delta_0T_B$, the zero-energy
state causes the anomalous Josephson effect in the $p_x$ symmetry
(See Ref.~\onlinecite{ya03-1}).
\end{thebibliography}
\end{document}